\documentclass{PoS}

\title{Hadrons in Nuclei}

\ShortTitle{Hadrons in Nuclei}

\author{\speaker{Laura Tolos}\\

          Institut f\"ur Theoretische Physik, University of Frankfurt, Max-von-Laue-Str. 1, 60438 Frankfurt am Main, Germany \\
          Frankfurt Institute for Advanced Studies,  University of Frankfurt, Ruth-Moufang-Str. 1, 60438 Frankfurt am Main, Germany \\
          Institute of Space Sciences (CSIC-IEEC), Campus Universitat Aut\`onoma de Barcelona, Carrer de Can Magrans, s/n, 08193 Cerdanyola del Vall\`es,
Spain 
\\
        E-mail: \email{tolos@th.physik.uni-frankfurt.de}}

\abstract{A review on the state-of-the-art of heavy hadrons in nuclei is presented. In particular, the properties of mesons with strangeness and charm  in  matter are discussed, paying a special attention to the formation of exotic bound states in nuclei.}

\FullConference{XVII International Conference on Hadron Spectroscopy and Structure - Hadron2017\\
		25-29 September, 2017\\
		University of Salamanca, Salamanca, Spain}

\begin{document}

\section{Introduction}

Over the last decades the properties of hadrons in nuclei have been matter of intense investigation. On the one hand, it is believed that hadrons in nuclei is an excellent scenario to test certain symmetries of the 
theory of the strong interaction, Quantum Chromodynamics (QCD), such as the chiral symmetry in the low-energy regime or heavy-quark symmetries as hadrons with charm or beauty content are produced in the laboratory.  On the other hand, it is of crucial importance to understand the excitation mechanisms in the nucleus as well as the nature of certain excited hadronic states, whose structure could be studied in a hot and/or dense nuclear medium.

In order to address hadrons in nuclei one can resort to theoretical and/or experimental analyses \cite{Metag:2017yuh,Rapp:2011zz}. From the theoretical side, there is an extensive variety of models that aim at understanding the properties of hadrons in nuclei, ranging from relativistic-mean field models (RMF), Nambu-Jona-Lasinio schemes (NJL), quark-meson coupling models (QMC), QCD sum-rule studies to unitarized approaches based on effective theories or meson-exchange models. Experimentally, photon-, electron-, neutrino- and hadron-induced reactions as well as heavy-ion collisions (HiCs) offer a gateway to the properties of hadrons in nuclei. The measurement of transparency ratios is extremely useful for the analysis of the imaginary part of the hadron-nucleon interaction in matter, whereas the excitation functions and the meson-momentum distributions are of fundamental importance to understand the real part of the interaction  \cite{Metag:2017yuh}. Moreover, it is crucial to connect the theoretical predictions to the experimental results, using transport model calculations or collision models based on nuclear spectral functions, to fully understand the dynamics of hadrons in nuclei.

In this paper we concentrate on the analysis of the properties of mesons with strangeness and charm content that interact with nuclei. Nowadays, strange and charmed  hadrons are being produced in nuclear and particle facilities, such as GSI, CERN or RHIC, while they are the subject of future experimental programs, such as FAIR, NICA or J-PARC.

\section{Strangeness in nuclei}

Strangeness in nuclei has received a lot attention in connection with the study of neutron stars interior \cite{Watts:2016uzu}, the properties of exotic atoms \cite{Friedman:2016rfd}, and strangeness production in heavy-ion collisions (HICs) \cite{Hartnack:2011cn}. In particular, the dynamics of strange mesons, such as $\bar K$, in vacuum and in the nuclear medium is still a challenge for theory and experiments.  In this section the $\bar K N$ interaction is studied, paying a special attention to the role of the $\Lambda(1405)$ and the formation of bound states, such as $\bar KNN$. Moreover, the production and propagation of strangeness in heavy-ion collisions (HICs) is investigated, in view of the present and forthcoming  experimental programs on strangeness.

\subsection{$\bar K N$ interaction: the  $\Lambda(1405)$}

The $\bar K N$ scattering in the $I=0$ channel is governed by the presence of the $\Lambda(1405)$, located only 27 MeV below the $\bar K N$ threshold. The dynamical origin of the $\Lambda(1405)$ dates back more than 50 years ago to the work of Dalitz and Tuan \cite{Dalitz:1959dn}. Recently, it has been revisited by means of  unitarized theories using meson-exchange models \cite{MuellerGroeling:1990cw,Haidenbauer:2010ch} or chiral Lagrangians \cite{Kaiser:1995eg, Oset:1997it, Oller:2000fj,Lutz:2001yb,GarciaRecio:2002td,Borasoy:2005ie,Oller:2006jw}, these latter ones analyzing the effects of including a complete basis of meson-baryon channels, studying the differences in the regularization of the equations, including s- and u-channel Born terms in the Lagrangian, implementing next-to-leading (NLO) contributions,... .  All these recent efforts have culminated in establishing the $\Lambda(1405)$ as a superposition of two poles of the scattering matrix \cite{Jido:2003cb}, that are generated dynamically from the unitarized coupled-channel scheme.

A renewed interest in the $\bar K N$ interaction has been developed in the past years after the availability of a more precise measurement of the energy shift and width of the $1s$ state in kaonic hydrogen by the SIDDHARTA Collaboration at DA$\Phi$NE \cite{Bazzi:2011zj}, that has helped to clarify the discrepancies between the KEK \cite{Iwasaki:1997wf,Ito:1998yi} and the DEAR \cite{Beer:2005qi,Cargnelli:2005cf} measurements. The obtained value of the energy shift is $\Delta E = 283 \pm 36 \pm 6$ eV with a width of $\Gamma=541 \pm 89 \pm 22$ eV, in good agreement with KEK results. Furthermore, the SIDDHARTA measurement has provided new constraints on the theoretical predictions reported in \cite{Ikeda:2011pi,Guo:2012vv,Mai:2012dt,Feijoo:2015yja}.

\begin{figure}[t]
\includegraphics[width=.5\textwidth]{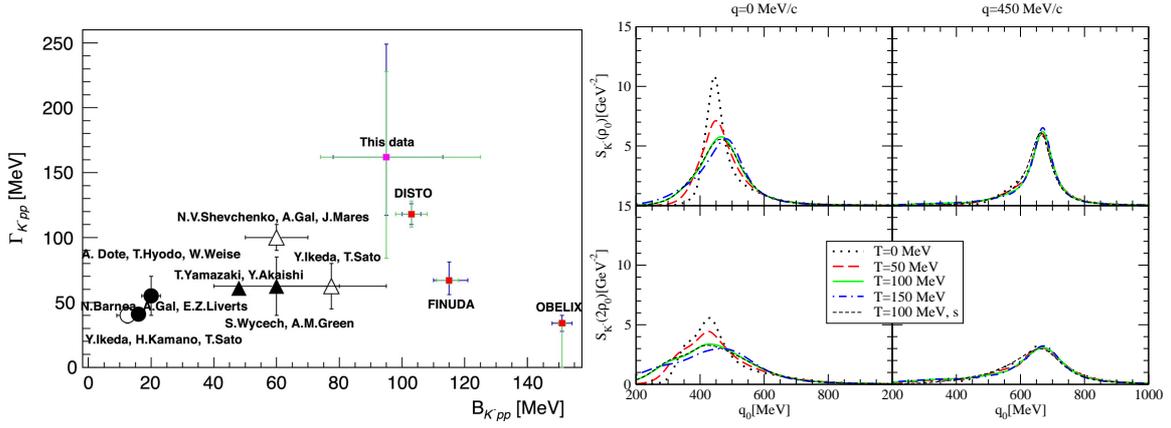}
\includegraphics[width=.5\textwidth]{Kbar.eps}
\caption{Left: Comparison of the binding energy and width of the $K^-pp$ between experiments and theoretical predictions, taken from \cite{Nagae:2016cbm}. Right: $\bar K$ spectral function for different densities, temperatures and momenta, taken from \cite{Tolos:2008di}.}
\label{fig1}
\end{figure}

\subsection{$\bar K NN$ state} 

The dynamical generation of the two-pole structure of the $\Lambda(1405)$ indicates that the $\bar K N$ interaction might be attractive enough to produce bound states. Indeed, it has been suggested that $\bar K$-nuclear clusters may form, such as the $\bar K NN$ in isospin $I=1/2$. The $I=1/2$ $\bar K NN$ state has been extensively studied, both theoretically and experimentally,  as shown in the left-hand side of {\bf Fig.~\ref{fig1}} (see Ref.~\cite{Nagae:2016cbm} and references therein).  This state was initially seen by the FINUDA \cite{Agnello:2005qj}, DISTO \cite{Yamazaki:2010mu} and OBELIX \cite{Bendiscioli:2009zz} Collaborations, but could be explained by means of conventional processes \cite{Ramos:2008zza} or not be reproduced  \cite{Agakishiev:2014dha}. Experiments performed by the Spring8/LEPS \cite{Tokiyasu:2013mwa}, J-PARC E15 \cite{Hashimoto:2014cri}, AMADEUS \cite{Doce:2015ust} Collaborations do not find any state, or, if found \cite{Nagae:2016cbm,Ichikawa:2014ydh}, may have other interpretation, such as as a possible $I=3/2$, $J^{\pi}=2^+$ resonance near the $\pi \Sigma N$ threshold  \cite{Garcilazo:2012rh}. More recently, the J-PARC E15 experiment has found a structure near the $\bar K NN$ threshold \cite{Sada:2016nkb}, that has been interpreted as a $\bar K NN$ bound state with a binding energy of $\sim 20-40$ MeV \cite{Sekihara:2016vyd}.

\subsection{$\bar K N$ in matter}

Antikaonic atoms \cite{Friedman:2016rfd} give us  information on the antikaon interaction with nucleons. The antikaon-nucleus potential has been extracted from best-fit analysis of antikaonic-atom data and some solutions agree with a very strongly attractive potential of the order of -200 MeV at normal saturation density $\rho_0$. However, some criticism has been raised because the antikaonic-atom data only tests matter at the surface of the nucleus. Recent analysis on $K^- N$ scattering amplitudes from chiral SU(3) effective field theories supplemented with phenomenological terms for $K^-$ multinucleon interactions indicate that antikaonic atoms are insensitive to densities above $\rho_0$ \cite{Friedman:2016rfd}.

Early works based on relativistic mean-field models \cite{Schaffner:1996kv} also obtained very deep potentials of a few hundreds of MeVs at $\rho_0$. However, later approaches on unitarized theories in coupled channels based on the chiral effective theory \cite{Lutz:1997wt,Ramos:1999ku} or on meson-exchange potentials \cite{Tolos:2000fj,Tolos:2002ud} obtain a potential much less attractive. In fact, in the unitarized coupled-channels models, the attraction is a consequence of the modified $s$-wave $\Lambda(1405)$ resonance in the medium due to Pauli blocking \cite{Koch:1994mj}, together with the self-consistent inclusion of the $\bar K$ self-energy \cite{Lutz:1997wt}, and the implementation of self-energies of the mesons and baryons in the intermediate states \cite{Ramos:1999ku}. As a result, the $\bar K$ spectral function can be obtained, as shown in the right-hand side of Fig.~\ref{fig1}. The $\bar K$ spectral function shows that $\bar K$ in matter feel a slight attraction while acquiring a remarkable width. Moreover, the knowledge of higher-partial waves beyond $s$-wave \cite{Tolos:2008di,Tolos:2006ny,Lutz:2007bh,Cabrera:2009qr,Cabrera:2014lca} becomes essential for analyzing the results of HiCs at beam energies below 2GeV per nucleon \cite{Cassing:2003vz,Tolos:2003qj}. 

\subsection{Strangeness production in HICs}

The production of $K$ and $\bar K$ close to threshold has been extensively investigated in low-energy HICs by the KaoS \cite{Forster:2007qk}, FOPI \cite{Lopez:2010mb} and HADES Collaborations \cite{Agakishiev:2014moo}. The analysis of experimental data together with microscopic transport approaches have permitted drawing several conclusions regarding the production mechanisms and the freeze-out conditions of strange mesons. However, the role of the in-medium properties of strange hadrons in their production and production in HiCs is still an open question. Recent results from HADES and FOPI indicate that, while the $K^+$ shows a repulsive interaction in matter, the $\Phi$ decay into $K^-$  washes out the effects of the $K^-$ potential in the spectra and flow \cite{leifels}. Therefore, more systematic and high statistic data on $K^-$ production are necessary, while further information in elementary reactions is required.

\section{Open charm in nuclei}

The interest in the properties of open and hidden charmed mesons was triggered in HiCs due to the possible charmonium suppression  as a probe for the formation of quark-gluon plasma (QGP). Nowadays, the nature of newly observed baryon and meson states with the charm degree of freedom is a matter of high interest in connection with many on-going experiments, such as BESIII, BelleII, ALICE, LHCb, amongst others, as well as with planned facilities, e.g. FAIR, NICA and the J-PARC upgrade.  The goal is to understand whether these states can be accommodated within the quark model picture and/or qualify better as being dynamically generated via hadron-hadron scattering processes. To this end, a large part of the experimental program in hadronic physics at PANDA (FAIR)  will be devoted to charmonium spectroscopy. Also, the CBM (FAIR) experiment will extend the GSI program for in-medium modification of hadrons in the light quark sector and provide the first insight into the charm-nucleus interaction. Indeed, the influence of medium modifications in the charmonium production at finite baryon densities would affect the formation of the QGP phase of QCD at high densities.

\subsection{ $D N$ interaction: the $\Lambda_c(2595)$}

Given the success of unitarized coupled-channel approaches in the description of some of the existing experimental data in the light-quark sector,  the charm degree of freedom has been recently incorporated in these models and several experimental states have been described as dynamically-generated baryon molecules  (see Ref.~\cite{Tolos:2013gta} and references therein).  This is the case, for example, of the $\Lambda_c(2595)$, which is the charmed counterpart of the $\Lambda(1405)$. 

Whereas a separable potential for the bare meson-baryon interaction with no strange degree of freedom was assumed in \cite{Tolos:2004yg}, later on unitarized approaches were based on a bare meson-baryon interaction saturated with the $t$-channel exchange of vector mesons between pseudoscalar mesons and baryons in the zero-range approximation \cite{Hofmann:2005sw,Mizutani:2006vq} or using the full $t$-dependence  \cite{JimenezTejero:2009vq}.  Other approaches have made use of the J\"ulich meson-exchange model \cite{Haidenbauer:2010ch,Haidenbauer:2007jq}, while some others have relied on the hidden gauge formalism \cite{Wu:2010jy, Oset:2012ap}. More recent schemes incorporate heavy-quark symmetry constraints explicitly,  such as those based on a pion-exchange model \cite{Yamaguchi:2013ty,Hosaka:2016ypm} or  on an extended Weinberg-Tomozawa  interaction for four flavors that includes pseudoscalar and vector mesons together with $1/2^+$ and $3/2^+$ baryons  \cite{GarciaRecio:2008dp, Gamermann:2010zz, Romanets:2012hm,Garcia-Recio:2013gaa}. In all these unitarized coupled-channel models, the $\Lambda_c(2595)$ is obtained dynamically, some of them \cite{Haidenbauer:2010ch,Hofmann:2005sw,GarciaRecio:2008dp, Gamermann:2010zz, Romanets:2012hm,Garcia-Recio:2013gaa} even obtaining a double-pole structure, in a similar manner as found for the $\Lambda(1405)$.

\subsection{ $\bar D N$ interaction}

The $C=-1$ has also been investigated within unitarized coupled-channel models, pion-exchange schemes with heavy-quark symmetry constraints or chiral quark models \cite{Hofmann:2005sw,Haidenbauer:2007jq,Yamaguchi:2013ty,Hosaka:2016ypm,Gamermann:2010zz,Carames:2012bd}.  Interestingly, some of the models find a $J=1/2$ state close to the $\bar DN$ threshold \cite{Yamaguchi:2013ty,Hosaka:2016ypm,Gamermann:2010zz}. In Ref.~\cite{Gamermann:2010zz} this state was generated  by the $ \bar D N$ and $\bar D^* N$ coupled channel dynamics, and it appears to be a consequence of treating heavy pseudoscalars and heavy vector mesons on an equal footing, because no resonance would be generated unless $\bar D^* N$ channel is considered.

\subsection{ $D NN$ and $\bar D NN$ states} 

Given the fact that the $DN$ and $\bar DN$ interactions are so attractive that allow for the formation of bound states, the question arises whether $D$ or $\bar D$-nuclear clusters may form.
In \cite{Bayar:2012dd} a $I=1/2$, $J=0^-$ $DNN$ state was found with mass $~3500$ MeV and width $\sim 20-40$ MeV, being interpreted as a quasibound state of the $\Lambda_c(2595)$ and a nucleon. Moreover, in  \cite{Yamaguchi:2013hsa} a state with $I=1/2$, $J=0^-$ and 5.2 MeV binding was found together with a $I=1/2$, $J=1^-$ state at 111.2 MeV above threshold.

\begin{figure}[t]
\begin{center}
\includegraphics[width=.3\textwidth,angle =-90]{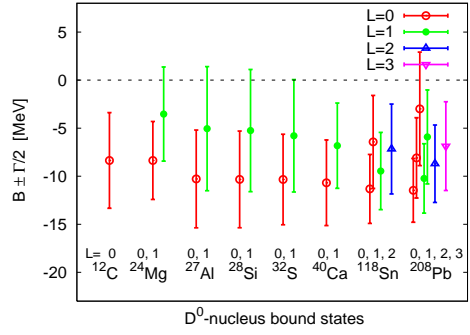}
\end{center}
\caption{$D^0$-nucleus bound states, taken from \cite{GarciaRecio:2010vt}. }
\label{figd0}
\end{figure}

\begin{figure}[t]
\begin{center}
\includegraphics[width=.45\textwidth]{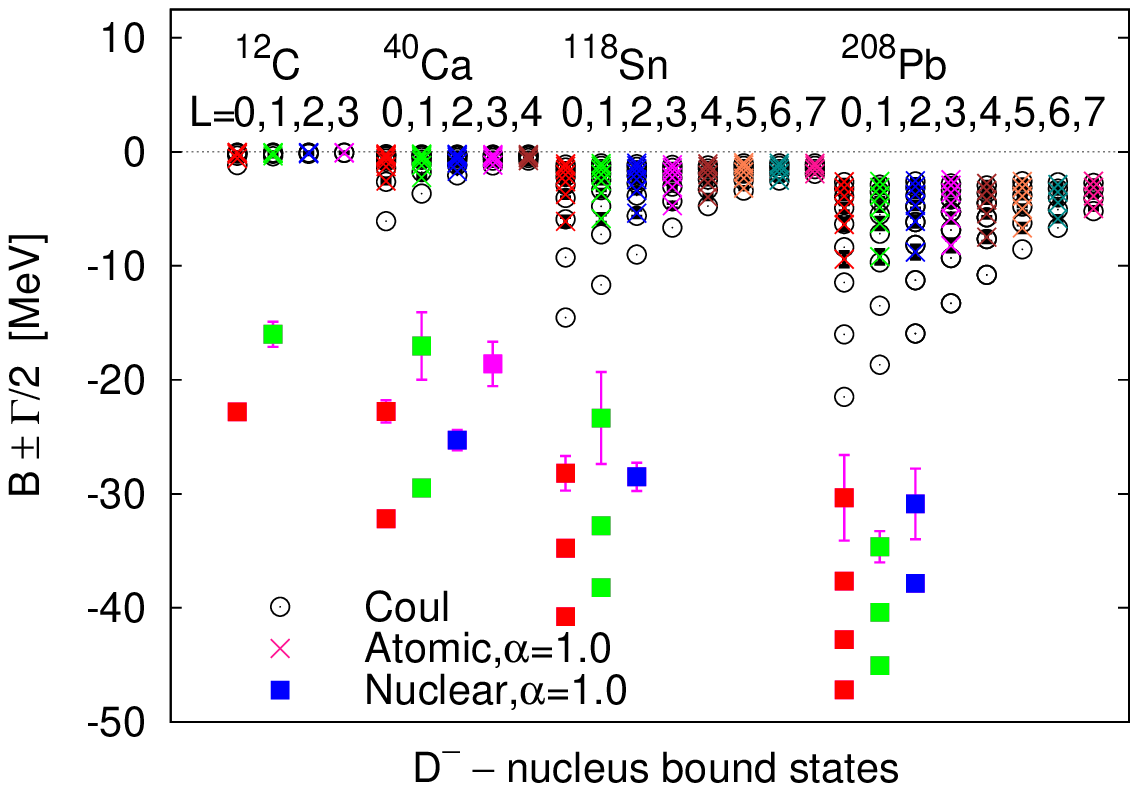}
\includegraphics[width=.45\textwidth]{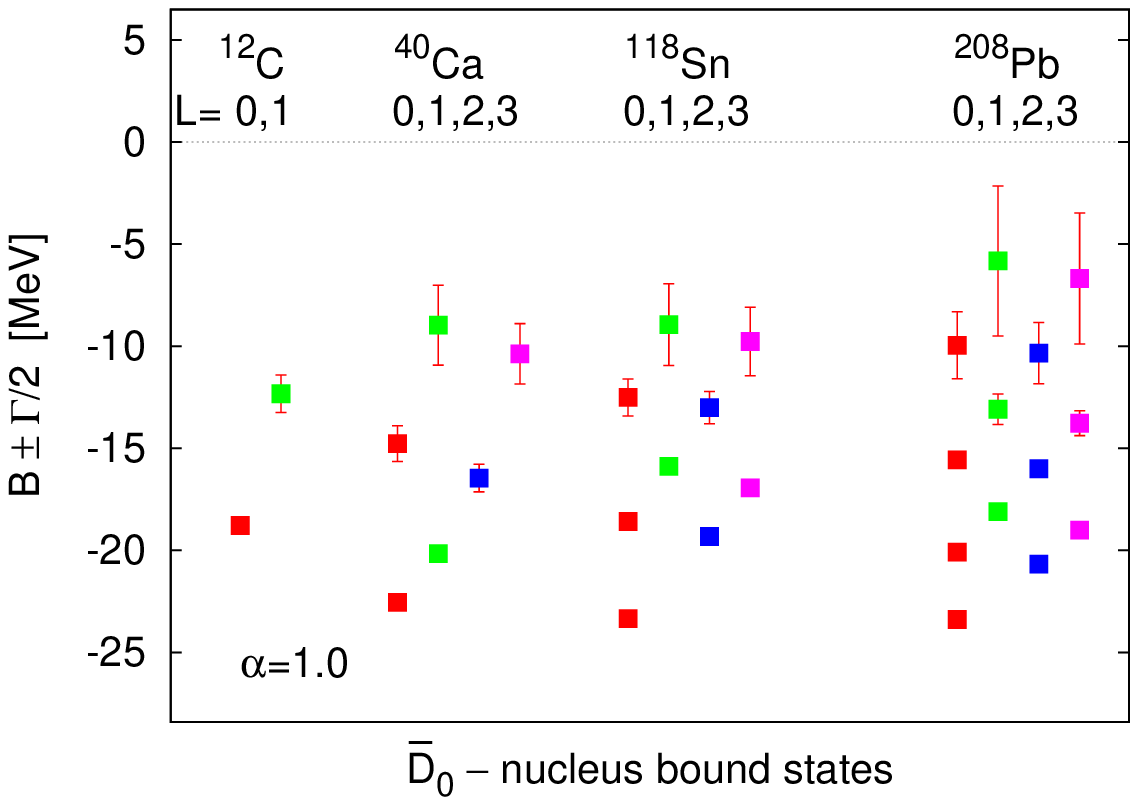}
\end{center}
\caption{$D^-$ and $\bar D^0$- nucleus bound states, taken from \cite{GarciaRecio:2011xt}.}
\label{figdm}
\end{figure}

\subsection{Open charm in matter}

The properties of open-charm mesons in matter have been object of theoretical interest due to the consequences for charmonium suppression, as observed at SPS energies by the NA50 collaboration. The change of the properties of D mesons in matter would modify the $J/\Psi$ absorption in a hot and dense nuclear medium and can provide an explanation for $J/\Psi$ suppression. Furthermore, there have been speculations about the existence of $D$-meson bound states in nuclei \cite{Tsushima:1998ru}. 

Several theoretical works have addressed the properties of open-charm mesons in dense baryonic matter:  QMC schemes \cite{Tsushima:1998ru}, QCD sum-rule approaches \cite{Hayashigaki:2000es,Hilger:2011cq, Suzuki:2015est}, NJL models \cite{Blaschke:2011yv}, chiral effective models in hot and dense matter \cite{Mishra:2003se} or pion-exchange approaches with heavy-quark symmetry constraints  \cite{Yasui:2012rw}. The full spectral features (mass and width) of the open-charm mesons in dense nuclear matter have been obtained in self-consistent unitarized coupled-channel schemes, where the intermediate meson-baryon propagators contain different sources of density dependence \cite{GarciaRecio:2011xt,Tolos:2004yg,Mizutani:2006vq,Tolos:2005ft,Lutz:2005vx,Tolos:2007vh,JimenezTejero:2011fc,Tolos:2009nn}.

\subsection{D-mesic nuclei}
A possible experimental scenario for the detection of the changes in matter of the properties of open-charm mesons would be the formation of D-mesic nuclei, where a D-meson binds in nuclear orbits.
In fact, $D$ and $\bar D$-meson bound states in $^{208}$Pb were predicted in Ref.~\cite{Tsushima:1998ru}, relying upon an attractive  $D$ and $\bar D$ -meson potential in the nuclear medium,  obtained 
within a QMC model. 

Within the unitarized coupled-channel model of Ref.~\cite{GarciaRecio:2010vt}, it is found that $D^0$-nucleus states are weakly bound (see Fig.~\ref{figd0}), in contrast to previous results using the QMC model \cite{Tsushima:1998ru},  while having significant widths. The best chances for observation of bound states are in the region of $^{24}\mbox{Mg}$, provided an orbital angular momentum separation can be done, where there is only one $s-$ bound state and its half width is about a factor of two smaller than the binding energy (see Fig.~\ref{figd0}). With regards to $D^+$-nuclear states, the Coulomb interaction prevents the formation of observable bound states.  As for  $\bar D$-mesic nuclei, not only $D^-$ but also $\bar{D}^0$ bind in nuclei as seen in Fig.~\ref{figdm}. The spectrum contains states of atomic and of nuclear types for all nuclei for $D^-$  while, as expected, only nuclear states are present for $\bar{D}^0$ in nuclei. Compared to the pure Coulomb levels, the atomic states are less bound. The nuclear ones are more bound and may present a sizable width, existing only for low angular momenta \cite{GarciaRecio:2011xt}.  This is in contrast to \cite{Tsushima:1998ru} for $^{208}$Pb, but close to results in \cite{Yasui:2012rw}. 

The experimental detection of $D$ and $\bar D$-meson bound states is, though, a difficult task \cite{Yamagata-Sekihara:2015ebw}. The formation of D-mesic nuclei with antiprotons beams at PANDA (FAIR) might be possible if ($\bar p$, D+N) or ($\bar p$, D+2N) reactions with small or even zero momentum transfer are produced, although the formation cross sections could be suppressed because of the complexity of the reaction mechanisms. Other competing mechanisms could involve the emission of pions by intermediate $D^*$ or $\bar D^*$ with subsequent trapping of the pseudoscalar charmed mesons by the final nucleus \cite{Yamagata-Sekihara:2015ebw}.

\section*{Acknowledgements}
L.T. acknowledges support from the Heisenberg Programme of the Deutsche Forschungsgemeinschaft under the Project Nr. 383452331,  the Ram\'on y Cajal research programme, FPA2013-43425-P and FPA2016-81114-P Grants from MINECO, and THOR COST Action CA15213.

\bibliographystyle{JHEP}
\bibliography{mybib}

\end{document}